\documentclass[a4paper,11pt]{article}
\usepackage{amsmath,amssymb}
\usepackage{graphicx,color}
\numberwithin{equation}{section}
\usepackage{cite}
\usepackage{bm}
\usepackage{dcolumn}
\newcommand{\be}{\begin{equation}}
\newcommand{\bea}{\begin{eqnarray}}
\newcommand{\eea}{\end{eqnarray}}
\newcommand{\ba}{\begin{array}}
\newcommand{\ea}{\end{array}}
\newcommand{\ee}{\end{equation}}

\expandafter\ifx\csname mathbbm\endcsname\relax

\else

\fi \textheight 22cm \textwidth 15cm \topmargin 1mm
\usepackage[T1]{fontenc} 

\usepackage{tikz}
\usetikzlibrary{arrows,snakes,backgrounds}
\usetikzlibrary{decorations.pathmorphing,backgrounds,shapes,arrows,shadows}
\tikzset{zigzag/.style={decorate,decoration=zigzag}}
\tikzset{snake it/.style={decorate, decoration=snake}}
\makeatletter
\def\@hex@@Hex#1%
 {\if a#1A\else \if b#1B\else \if c#1C\else \if d#1D\else
  \if e#1E\else \if f#1F\else #1\fi\fi\fi\fi\fi\fi \@hex@Hex}
\makeatother

\usepackage[all]{xy}
\usepackage[percent]{overpic}
\usepackage{slashed}
\usepackage{wrapfig}
\usepackage{tabu}
\usepackage{diagbox}
\usepackage{mathrsfs,amsmath,amssymb,amsthm,amsfonts,tikz,graphicx,accents,hyperref, color}
\usepackage{dsfont,epiolmec, latexsym, stmaryrd, comment}
\usepackage{slashed,ccaption}
\usepackage{mathrsfs, calligra}
\usepackage{leftidx}
\usepackage{import}
\usepackage{multirow}
\usepackage{amsfonts}
\usepackage{pifont}
\usepackage{tabularx}
\usepackage{cancel}
\usepackage[utf8]{inputenc}
\usetikzlibrary{intersections,calc}
\usepackage{ifthen}
\usepackage{amsmath}
\usepackage{cancel}
\usepackage{caption}

\usepackage{array}
\hypersetup{ linktoc=all,
    colorlinks, linkcolor={palatinateblue},
    citecolor={brightpink}, urlcolor={amaranth}
}

\graphicspath{{Images/}}

\renewcommand{\d}[1]{\ensuremath{\operatorname{d}\!{#1}}}

\def\sideremark#1{\ifvmode\leavevmode\fi\vadjust{\vbox to0pt{\vss
 \hbox to 0pt{\hskip\hsize\hskip1em
 \vbox{\hsize2cm\tiny\raggedright\pretolerance10000
 \noindent #1\hfill}\hss}\vbox to8pt{\vfil}\vss}}}%
\DeclareSymbolFont{extraup}{U}{zavm}{m}{n}
\DeclareMathSymbol{\varheart}{\mathalpha}{extraup}{86}
\DeclareMathSymbol{\vardiamond}{\mathalpha}{extraup}{87}
\makeatletter
\renewcommand*{\@fnsymbol}[1]{\ensuremath{\ifcase#1\or \clubsuit \or \vardiamond \or \varheart\or
    \spadesuit\or \mathparagraph\or \|\or **\or \dagger\dagger
    \or \ddagger\ddagger \else\@ctrerr\fi}}
\makeatother

\definecolor{rosy}{RGB}{230,235,252}
\definecolor{myframetitle}{RGB}{90,89,170}
\definecolor{myblocktitle}{RGB}{140,185,249}
\definecolor{mytitle}{RGB}{10,80,26}

\definecolor{darkgreen}{RGB}{27,130,45}
\definecolor{darkblue}{rgb}{0,0,0.3}
\definecolor{darkred}{rgb}{0.7,0,0}

\definecolor{light gray}{RGB}{220,220,220}
\definecolor{dark purple}{RGB}{108,0,217}
\definecolor{pink}{RGB}{190,20,100}
\definecolor{orang}{RGB}{193,63,0}
\definecolor{green}{RGB}{11,98,17}
\definecolor{darkpink}{RGB}{153,0,76}
\definecolor{bluegreen}{RGB}{0,102,102}
\definecolor{greenlagan}{RGB}{0,102,0}
\definecolor{redgreen}{RGB}{102,102,0}
\definecolor{Redgreen}{RGB}{153,76,0}
\definecolor{vividviolet}{rgb}{0.62, 0.0, 1.0}
\definecolor{amaranth}{rgb}{0.9, 0.17, 0.31}
\definecolor{palatinateblue}{rgb}{0.15, 0.23, 0.89}
\definecolor{brightpink}{rgb}{1.0, 0.0, 0.5}
\definecolor{cornflowerblue}{rgb}{0.39, 0.58, 0.93}
\definecolor{deepcarminepink}{rgb}{0.94, 0.19, 0.22}
\definecolor{radicalred}{rgb}{1.0, 0.21, 0.37}

\newcommand\bc[1]{\boldsymbol{\mathcal{#1}}}

\DeclareFontFamily{OT1}{rsfs}{}

\DeclareFontShape{OT1}{rsfs}{m}{n}{ <-7> rsfs5 <7-10> rsfs7 <10->rsfs10}{} 

\DeclareMathAlphabet{\mycal}{OT1}{rsfs}{m}{n}
\makeatletter \@addtoreset{equation}{section}

\newcommand\tcb{\textcolor{blue}}

\begin{document}
\begin{titlepage}
\begin{center}
{\Large {\bf  Null Surface Thermodynamics}\\}
\vspace*{15mm}
\vspace*{1mm}
{H.~Adami$^{a,b}$ , M.M.~Sheikh-Jabbari$^{c}$, V.~Taghiloo$^{c,d}$, H.~Yavartanoo$^{b}$ }

 \vspace*{1cm}

{\it ${}^a$ Yau Mathematical Sciences Center, Tsinghua University, Beijing 100084, China\\}
 \vspace*{0.5cm}
{\it ${}^b$ Beijing Institute of Mathematical Sciences and Applications (BIMSA), Huairou District, \\ Beijing 101408, P. R. China \\}
\vspace*{0.5cm}
{\it ${}^c$ School of Physics, Institute for Research in Fundamental
Sciences (IPM), \\ P.O.Box 19395-5531, Tehran, Iran\\}
\vspace*{0.5cm}
{\it ${}^d$ Department of Physics, Institute for Advanced Studies in Basic Sciences (IASBS), \\
P.O. Box 45137-66731, Zanjan, Iran \\}
\vspace*{0.5cm}

E-mails: hamed.adami@bimsa.cn, 
jabbari@theory.ipm.ac.ir,  v.taghiloo@iasbs.ac.ir, yavar@bimsa.cn

\vspace*{2cm}
\end{center}

\begin{abstract}

We establish  that  boundary degrees of freedom associated with a generic co-dimension one null surface in  $D$ dimensional pure Einstein gravity naturally admit a thermodynamical description. We expect the \textit{null surface thermodynamics}  to universally follow as a result of the diffeomorphism invariance of the theory, not relying on other special features of the null surface or the gravity theory. Using standard surface charge analysis and covariant phase space method, we formulate  laws of null surface thermodynamics which are local equations over an arbitrary null surface {paralleling local versions the zeroth and first laws and the Gibbs-Duhem equation}. This thermodynamical system is generally an open system and can be closed only when there is no flux of gravitons through the null surface. Our analysis  extends the usual black hole thermodynamics to a universal feature of any area element on a generic null surface. We discuss the relevance of our study for the membrane paradigm and black hole microstates. 
\end{abstract}
\end{titlepage}

\section{Introduction}
{Despite apparent differences,} there are various hints that gravity, as formulated by Einstein's General Relativity (GR), and thermodynamics are closely related to each other, both at conceptual and formulation levels. Perhaps, the first hint was already in the statement of Einstein's equivalence principle and  the universality of   GR and thermodynamics. In the context black hole physics the resemblance between laws of black hole mechanics and laws of thermodynamics \cite{Bardeen:1973gs} was gradually completed into the equivalence of the two \cite{Bekenstein:1973ur, Bekenstein:1974ax, Hawking:1974rv, Hawking:1975sw , Jacobson:1997se, Wald:1999vt, Padmanabhan:2003gd,  Wald:2018wro, Wall:2018ydq}.

The connection is not limited to black holes. In a seminal paper \cite{Unruh:1976db}, Unruh showed that there is a non-zero temperature associated with a generic accelerated observer, as required  by the equivalence principle.  The next remarkable step was provided by Wald who showed that black hole entropy is a conserved charge associated with bifurcate Killing horizons \cite{Wald:1993nt} and derived the first law of thermodynamics for generic probes around such black holes as a direct consequence of diffeomorphism invariance \cite{Iyer:1994ys}. And finally, Jacobson derived Einstein's field equations  from the first law of thermodynamics adapted around a null surface \cite{Jacobson:1995ab}, see also \cite{Ashtekar:2002ag,Hayward:2004dv} and \cite{Paranjape:2006ca, Padmanabhan:2007en, Padmanabhan:2009vy}. The connection between gravity and thermodynamics was also reinforced through the holographic principle \cite{Bousso:2002ju} and the AdS/CFT duality \cite{Aharony:1999ti} and presented  bluntly in \cite{Verlinde:2010hp}.

{Thermodynamics aspects of  black holes is generically attributed to black hole microstates and to the horizon. Horizon is typically a null surface which is the boundary of the spacetime as viewed by non-free fall observers outside the hole. Presence of boundaries lead to existence of `boundary degrees of freedom' which reside only at the (timelike or null) boundaries and interact with bulk (graviton) modes. Black hole microstates may be sought for among these boundary degrees of freedom.}
With this motivation, we study gravity theory on spacetimes with a null boundary. This boundary  can be an arbitrary one in spacetime {and is not necessarily horizon of a black hole}. This has been the research program pursued in some recent {works} \cite{Adami:2020amw, Grumiller:2020vvv, Adami:2020ugu, Adami:2021sko} and in particular in \cite{Adami:2021AAA, Grumiller:2019fmp, Donnay:2015abr, Donnay:2016ejv, Afshar:2016wfy, Afshar:2016uax,Afshar:2016kjj, Hopfmuller:2016scf, Hopfmuller:2018fni, Donnay:2018ckb, Chandrasekaran:2018aop, Chandrasekaran:2019ewn, Chandrasekaran:2020wwn, Ciambelli:2021vnn,Freidel:2021fxf, Freidel:2021cbc}. It has been established that {(see \cite{Grumiller:2020vvv,Adami:2021AAA} and references therein)} the most general solution phase space of $D$ dimensional pure Einstein gravity theory around a given null surface besides the bulk graviton modes involves boundary modes parametrized by $D$ arbitrary functions over the $D-1$ dimensional null boundary.

{In this work we {show that the thermodynamics description is not limited to (black hole) horizons. We recast the equations used in boundary symmetry and charge analysis for a pure $D$ dimensional Einstein gravity in presence of a null boundary as the local first law of thermodynamics and the Gibbs-Duhem equation. We construct the solution phase space governing the boundary degrees of freedom and show it can be naturally viewed as an open  thermodynamic phase space. 

This open thermodynamics system} can be closed if we turn off the graviton flux passing through the null surface. The latter together with an extra relation among the chemical potentials and associated surface charges \eqref{zeroth-law-H}, yields the statement of the local zeroth law. These relations are universal and independent of details of gravitational theory or the null surface. Our derivation only relies on diffeomorphism invariance of the theory and we expect our thermodynamical description to be true for any generally invariant theory of gravity.}

\section{Null Surface Solution Phase Space, A Review}\label{sec:review}

We start with a $D$ dimensional ($D\geq 3$)  generic metric adopting $v,r, x^A$ coordinates, 
\begin{equation}\label{G-F-M-01}
    \d s^2=  -V \d v^2 + 2 \eta \d v \d r + g_{_{AB}} \left( \d x^A + U^A \d v\right) \left( \d x^B + U^B \d v\right)\,, 
\end{equation}
such that $r=0$ is the null surface $\mathcal{N}$ and metric coefficients admit the near  ${\cal N}$ expansion,
\begin{equation}\label{nearN-expansion}
    \begin{split}
        &V=-\eta \left(  {\Gamma - \frac{2}{D-2} \frac{\mathcal{D}_v \Omega}{\Omega}}+ \frac{\mathcal{D}_v \eta}{\eta} \right)  r  + \mathcal{O}(r^2) \\
        &U^A= {\cal U}^A - r \frac{\eta}{\Omega} {\cal J}^A + \mathcal{O}(r^2) \\
        &g_{_{AB}}=\Omega_{AB} + \mathcal{O}(r) 
    \end{split}
\end{equation}
where all the fields are functions of $v,x^A$ and 
\be\label{OmegaAB-gammaAB}
\Omega_{AB}=\Omega^{2/(D-2)}\gamma_{{AB}}, \qquad \Omega:=\sqrt{\det \Omega_{AB}},\qquad  \det \gamma_{_{AB}} =1.
\ee
We use the definition 
\be
\mathcal{D}_v := \partial_v - \mathcal{L}_{\mathcal{U}}, 
\ee 
where $\mathcal{L}_{\mathcal{U}}$ is {the} Lie derivative along $\mathcal{U}^A$ {direction}. 
Let $\Theta$ be {the} expansion of vector field  generating the null surface $\mathcal{N}$ and $N_{_{AB}}$ be the \textit{news tensor} associated with flux of gravitons through ${\cal N}$:
\be\label{expansion-news}
\Theta :=\mathcal{D}_v \ln{\Omega}, \qquad N_{_{AB}}:= \frac{1}{2}\Omega^{2/(D-2)} \mathcal{D}_v \gamma_{_{AB}}
\ee
We use $\Omega^{AB}$ and $\Omega_{AB}$ respectively for raising and lowering capital Latin indices. Note that $N_{_{AB}}$ as defined above is {a} symmetric-traceless {tensor}.

The coefficients appearing in the metric are subject to Einstein field equations. In particular,  
there are Raychaudhuri and Damour equations which  play a crucial role in our analysis, see  \cite{Adami:2021AAA} for a more detailed treatment.  These two equations respectively are
\begin{subequations}\label{RD-EoM}
\begin{align}
    &{\cal D}_v  \Theta  + \frac{1}{2} \left( {\Gamma} - \frac{\mathcal{D}_v \eta }{\eta} \right)  \Theta +N_{AB}N^{AB}=0,\label{EoM-Raychaudhuri}\\
        &{\cal D}_v {\cal J}_{A}+\Omega\Theta \frac{\partial_A\eta}{\eta}+\Omega\partial_{A}\left( {\Gamma  - 2\Theta}  \right)+2 \Omega\bar{\nabla}^{B}N_{A B}=0.\label{EoM-Damour}
\end{align}
\end{subequations}
Here $\bar{\nabla}_{A}$ denotes {the} covariant derivative with respect to the metric $\Omega_{AB}$. The new variable ${\cal P}$ defined as
\be\label{P-def}
{{\cal P}:=\ln\frac{\eta}{\Theta ^2}}, 
\ee
may substitute $\eta$ and in terms of which  \eqref{RD-EoM} simplify to \begin{subequations}\label{RD-EoM-2}
\begin{align}
    &{\cal D}_v \Omega=\Omega\Theta,\label{Theta-Omega}\\
    &{\cal D}_v {\cal P} =\Gamma +\frac{2}{\Theta }N_{AB}N^{AB}  ,\label{EoM-cal-P}\\
        &{\cal D}_v {\cal J}_{A}+\Omega\Theta \partial_A{\cal P}+\Omega\partial_{A}{\Gamma}+2 \Omega\bar{\nabla}^{B}N_{A B}=0.\label{EoM-Damour-JP}\end{align}
\end{subequations}

\paragraph{Off-shell pre-symplectic form.} Starting from the Einstein-Hilbert action,
\be\label{EH-action}
S_{\text{\tiny EH}}=\frac{1}{16\pi G}\int \d{}r\d{}v \d{}^{D-2}x\ \eta\sqrt{\det{g_{_{AB}}}}\ L_{\text{\tiny EH}},\qquad L_{\text{\tiny EH}}=R-2\Lambda,
\ee
one can compute the usual Lee-Wald pre-symplectic form \cite{Lee:1990nz}  over the  set of geometries \eqref{G-F-M-01}, yielding\begin{equation}\label{presymplectic}
    \begin{split}
     \Omega_{\text{\tiny LW}}
             =  \frac{1}{16 \pi G} \int_{\mathcal{N}} \d v \d{}^{D-2} x \bigr[  \delta \mathcal{U}^{A} \wedge\delta {\cal J}_{A}  - \delta \Gamma\wedge \delta \Omega 
        +\delta ( \Omega \Theta ) \wedge \delta \mathcal{P}+ \delta\Omega_{AB}\wedge \delta (\Omega N^{AB}) \bigr].
    \end{split}
\end{equation}

While the above {expression} clearly shows which variables are canonical conjugate of each other, the functions appearing {there} are subject to {equations} \eqref{RD-EoM-2} and not all of them are independent. In other words, the solution phase space is obtained after imposing the constraints \eqref{RD-EoM-2} upon the parameter space and the symplectic form, e.g. using the Dirac bracket method or going to the reduced phase space. 

\paragraph{Null boundary symmetry generators.} 
The vector field
\begin{equation}\label{null-boundary-sym-gen}
    \xi =T\, \partial_v + r \left(\mathcal{D}_v T -  W  \right) \partial_r + \left( Y^A -r \eta  \partial^A T\right)\partial_A + \mathcal{O}(r^2)\, ,
\end{equation}
preserves the form of metric \eqref{G-F-M-01}, keeps $r=0$ a null surface and generates the following variations over the solution phase space functions
\begin{subequations}\label{delta-fields}
    \begin{align}
                     \delta_\xi \eta &=  T \partial_{v} \eta +2 \eta {\cal D}_v T-W\eta  +Y^A \partial_A \eta\,,\\ 
                      \delta_{\xi}\Gamma &=-{\cal D}_{v}(W-\Gamma T)+(Y^{A}+\mathcal{U}^A T)\partial_{A}\Gamma\,,\label{delta-Gamma} \\ 
                      \delta_\xi (\Omega\Theta) & 
                      = {\cal D}_v (T\Omega\Theta) + \mathcal{L}_{(Y+T\mathcal{U})}(\Omega\Theta)\,,\label{delta-DvOmega}\\  
                \delta_\xi \mathcal{U}^{A} &={\cal D}_v (Y^A + T \mathcal{U}^{A})\,,\label{delta-UA}\\
                  \delta_\xi \Omega_{AB} &=  2 T N_{AB}+\mathcal{L}_{(Y+T\mathcal{U})} \Omega_{AB} +\frac{2}{D-2}\, T \Theta  \Omega_{AB} \label{delta-OmegaAB}\, .
   \end{align}
\end{subequations}       
where ${\cal L}_Y$ denote the Lie derivative along $Y^A$, and for associated conjugate charges (see below) 
\begin{subequations}\label{delta-charges}
    \begin{align}      
        \delta_\xi\Omega &=T \Omega \Theta+ \mathcal{L}_{(Y+T \mathcal{U})}\Omega\, ,\label{delta-Omega}\\ 
        \delta_\xi {\cal P} &=T{\cal D}_v {\cal P}+\mathcal{L}_{(Y+T\mathcal{U})}{\cal P}-W %
        \, ,\label{delta-P}\\
      \delta_{\xi}{\cal J}_{A}&= T {\cal D}_v {\cal J}_A+\mathcal{L}_{(Y+T \mathcal{U})}{\cal J}_{A}+\Omega \big[\partial_{A}W- {\Gamma}\partial_{A}T-{2}{N_{A B}}\partial^{B}T\big]
    \,,\label{delta-JA}\\
              \delta_{\xi}N_{AB} &=\mathcal{D}_{v}(TN_{AB})+\mathcal{L}_{(Y+T\mathcal{U})}N_{AB}\,.\label{delta-NAB}
\end{align}
\end{subequations}
\paragraph{Surface charge variation.} One may compute the charge variation associated with the boundary symmetry generators using covariant phase space method  \cite{Lee:1990nz, Iyer:1994ys}. Detailed analysis yields \cite{Adami:2021AAA}
\begin{equation}\label{surface-charge-01}
\begin{split}
        \slashed{\delta} Q_{\xi}= \frac{1}{16\pi G} \int_{{\cal N}_v} \d{}^{D-2} x\bigg[ &\left(W-\Gamma T \right)\delta\Omega +(Y^{A}+{\cal U}^A T)\delta{\cal J}_{A} 
        +  T\Omega\Theta \delta{\cal P}  - T\Omega  \Omega^{AB} \delta N_{AB} \bigg] \, ,
\end{split}
\end{equation}
where ${\cal N}_v$ is a constant $v$ section on ${\cal N}$. This charge variation is an integral over $\sum_{i=1}^4 {\cal G}_i\ \delta \bc{Q}_i$, 
where $\bc{Q}_i$ parameterize the solution phase space. Among the four families, $N_{AB}$ corresponds to the bulk degrees of freedom while three others $\Omega, {\cal J}_A, {\cal P}$ parameterize boundary information. $\Gamma, {\cal U}^A$ functions which appear in  ${\cal G}_i$ are subject to field equations \eqref{RD-EoM-2} and  $\delta \bc{Q}_i$ subject to linearized equations of motion. 

The ${\cal G}_i$ are `field dependent' linear combinations of symmetry generators $T, W, Y^A$, notably ${\cal G}_i$ depend on $\Gamma, {\cal U}^A$ as well as $\Omega\Theta $ and $\Omega_{AB}$ and $\delta_\xi {\cal G}_i\neq 0$. The charge variation $\slashed{\delta} Q_{\xi}$, as stressed in the notation $\slashed{\delta}$, is hence not integrable.  $\Gamma, {\cal U}^A$ may be respectively solved for in terms of the charges using \eqref{EoM-cal-P} and \eqref{EoM-Damour-JP} and therefore all these coefficients may be represented through the charges. Note also that there are three symmetry generators and four towers of charges and these are  functions over the null surface ${\cal N}$. We crucially note that $\delta \Omega, \delta {\cal J}_A, \delta{\cal P}, \delta N_{AB}$ denote generic variations around solutions of {equations of motion} (EoM) and are subject to linearized field equations. These linearized equations may be viewed as equations for variations $\delta\Gamma, \delta{\cal U}^A$. The solution phase space is hence parametrized by the four tower of charges and their variations.

We close this part by giving expressions for three `zero mode' charges, $\xi=-r\partial_r, \xi=\partial_A$ and $\xi=\partial_v$. One may readily observe that the first two are integrable and the latter is not:
\begin{equation}\label{Charges-zero-mode-full}
\begin{split}
        Q_{-r\partial_r} &:= \frac{\mathbf{S}}{4\pi}=\frac{1}{16\pi G} \int_{{\cal N}_v} \d{}^{D-2} x\ \Omega,     \\ 
        Q_{\partial_A} &:= \mathbf{J}_A=\frac{1}{16\pi G} \int_{{\cal N}_v} \d{}^{D-2} x\ {\cal J}_A,\\
       \slashed{\delta} Q_{\partial_v} &:= \slashed{\delta}\mathbf{H}=\frac{1}{16\pi G} \int_{{\cal N}_v} \d{}^{D-2} x\ \left(-\Gamma \delta\Omega +{\cal U}^A \delta{\cal J}_{A}+\Omega \Theta  \delta {\cal P}-\Omega \Omega^{AB}\delta N_{AB} \right) \,.
\end{split}
\end{equation}
\paragraph{Surface charges and  flux in thermodynamics slicing.}
The charge variation may be split into Noether (integrable) part $Q^{\text{\tiny N}}$ and the `flux' part $F$: $\slashed{\delta} Q_{\xi}= \delta Q^{\text{\tiny N}}_\xi + {F}_\xi (\delta g ; g)$. $Q^{\text{\tiny N}}$ may be computed for the Einstein-Hilbert action using the standard Noether procedure, yielding
\begin{equation}\label{Integrable-part-Charge-Geometric}
    Q^{\text{\tiny N}}_\xi = \frac{1}{16\pi G} \int_{{\cal N}_v} \d{}^{D-2} x \left[  W\, \Omega+Y^{A}\, {\cal J}_{A}+ T\, \left( -\Gamma \Omega + \mathcal{U}^{A}{\cal J}_{A}\right) \right] \, ,
\end{equation}
and non-integrable flux part
\begin{equation}\label{Non-Integrable-part-Charge-Geometric}
F_{\xi}(\delta g ; g)= \frac{1}{16\pi G} \int_{{\cal N}_v} \d{}^{D-2} x \, T \left(  \Omega \delta \Gamma -{\cal J}_{A}\delta \mathcal{U}^{A} +\Omega \Theta \delta \mathcal{P}{-\Omega \Omega^{AB}\delta N_{AB}} \right) .
\end{equation}
Here we are assuming symmetry generators $T, W, Y^A$ to be field-independent, i.e. $\delta T=\delta W=0=\delta Y^A$. 

For later use, we also present the expressions for the zero mode Noether charges, 
\begin{equation}\label{Charges-zero-mode-Integrable}
\begin{split}
        Q^{\text{\tiny N}}_{-r\partial_r} &= \frac{1}{16\pi G} \int_{{\cal N}_v} \d{}^{D-2} x\ \Omega,     \\ 
        Q^{\text{\tiny N}}_{\partial_A} & =\frac{1}{16\pi G} \int_{{\cal N}_v} \d{}^{D-2} x\ {\cal J}_A,\\
        Q^{\text{\tiny N}}_{\partial_v} &:= \mathbf{E}=\frac{1}{16\pi G} \int_{{\cal N}_v} \d{}^{D-2} x\ (-\Gamma \Omega +{\cal U}^A {\cal J}_{A}  ) \, ,
\end{split}
\end{equation}
\paragraph{Balance or  `generalized charge conservation' equation.} In our general setup charges and the flux are given by integrals over co-dimension two surface ${\cal N}_v$. They are hence functions of `lightcone time' coordinate $v$ and the charges are not conserved. From the expressions above one can deduce,
\be\label{GCCE}
\frac{\d {} }{\d v} Q^{\text{\tiny N}}_{\xi}\approx -F_{\partial_v}(\delta_\xi g ; g),
\ee
where $\approx$ denotes on-shell equality. Eq.\eqref{GCCE}  may be viewed as (1) manifestation of the boundary EoM written in terms of charges; (2) 
a `generalized charge conservation equation' as it relates time dependence, or non-conservation, of the charge (as viewed by the null boundary observer) to the flux passing through the boundary; (3) how the passage of flux through the null boundary is `balanced' by the rearrangements in the charges. In this respect, it is very similar to the usual balance equation used at asymptotic null surfaces, which is now written for an arbitrary null surface in the bulk. Note also that the third viewpoint yield \textit{null surface memory effects} discussed in \cite{Adami:2021AAA}.

\section{Null Surface Thermodynamics}\label{sec:thermodynamics}

Consider a usual thermodynamical system with chemical potentials $\mu_i (i=1,2,\cdots, N$) and temperature $T$. This system is specified with charges $Q_i$, the entropy $S$ and the energy $E$; that is, there are $N+2$ charges and $N+1$ chemical potentials. The distinction between charges and associated chemical potentials is by convention and is specified with/specifies the ensemble. In microcanonical ensemble (which we have already assumed), the first law takes the form
\be\label{first-law}
\d E=T \d S+\sum_{i=1}^N\mu_i \d Q_i.
\ee
This equation implies that the LHS is an exact one-form over the thermodynamic space. Moreover,  chemical potentials and the charges are related by the Gibbs-Duhem relation
\be\label{GD-thermo}
S \d T+ \sum_{i=1}^N Q_i  \d \mu_i =0.
\ee
Together with the first law \eqref{first-law} this yields $E=TS+\sum_i\mu_i Q_i$. This equation  relates $E$ to the other charges and chemical potentials, e.g. $E=E(S,Q_i)$ (in microcanonical description) or $E=E(T, \mu_i)$ (in grandcanonical description). Depending on the ensemble chosen, $N+1$ number of chemical potentials and/or charges may be taken to be `independent' variables parameterizing the thermodynamical configuration space  and the rest of $N+1$ of them as functions of the former $N+1$ variables. In other words, the thermodynamic configuration space is $(N+1)$ dimensional and the change of ensemble is basically a canonical 
transformation 
the generator of which is the difference between various `energy' functions associated with each ensemble. 

\subsection{Null Boundary Thermodynamical Phase Space}\label{sec:Thermodynamical phase space}

Staring at the expression of the charge variation \eqref{surface-charge-01}, one can recognize that  functions parameterizing the solution space come in two categories: the bulk modes $N_{AB}$ (and its conjugate `chemical potential' determinant-free part of ${\Omega^{AB}}, \gamma^{AB}$) and the boundary modes. The latter may also be separated into those whose variation appears $\Omega, {\cal P}$  and ${\cal J}_A$, and those which appear only in the coefficients, in chemical potentials  $\Gamma, {\cal U}^A$. There are hence $D=1+1+(D-2)$ charges and $D-1=1+(D-2)$ chemical potentials.

We crucially note that if we treat $\Gamma, {\cal U}^A$ and associated charges $\Omega, {\cal J}_A$ as independent variables, ${\cal P}$ is special as it does not appear in the integrable part of the charges \eqref{Integrable-part-Charge-Geometric} and only appears in the expression for the flux \eqref{Non-Integrable-part-Charge-Geometric} through $\Omega\Theta \delta{\cal P}$ term. Moreover, as already remarked (\emph{cf.} discussions below \eqref{surface-charge-01}), the chemical potentials may be expressed in terms of the charges using field equations. Again we note at $\Theta=0$, ${\cal P}$ dependence completely drops out of the analysis.

Given all the above we are led to the following picture for the generic case. 
\begin{enumerate}
\item[I.] Null boundary solution space relevant to the null boundary thermodynamics consists of three parts: 
\begin{enumerate}
    \item[I.1)] $(D-1)$ dimensional `thermodynamic sector' parametrized by $(\Gamma, {\cal U}^A)$ and conjugate charges $(\Omega, {\cal J}_A)$;
    \item[I.2)] ${\cal P}$, which   only appears in  the flux \eqref{Non-Integrable-part-Charge-Geometric} and not in the Noether charge \eqref{Integrable-part-Charge-Geometric};
    \item[I.3)] the bulk mode parameterized by determinant free part of $\Omega^{AB}$ and its `conjugate charge' $N_{AB}$ which appear in the flux \eqref{Non-Integrable-part-Charge-Geometric}.
\end{enumerate}
    \item[II.] $N_{AB}$ parameterizes effects of the bulk and how they take the boundary system  out-of-thermal-equilibrium (OTE) whereas ${\cal P}$  parameterizes OTE within the boundary dynamics. Put differently,  OTE  may come from inner boundary dynamics and/or from the gravity-waves passing through the null boundary, parameterized by $N_{AB}$. 
     
    \item[III.] Expansion parameter $\Theta $ is a measure of OTE, from both bulk and boundary viewpoints. When $\Theta =0$ the system is completely specified by the $D-1$ dimensional thermodynamic phase space. 
    \item[IV.] The rest of the in-falling graviton modes parameterized through ${\cal O}(r)$ terms in $g_{_{AB}}$, do not enter in the boundary/thermo dynamics, as of course expected from usual causality and that the boundary is a null surface.
\end{enumerate}

We start with {the} local first law, then local Gibbs-Duhem equation and come to local zeroth law, specifying the subsectors which can be brought to a (local) equilibrium. Before moving on, we introduce a piece of useful notation. By $\boldsymbol{\mathcal{X}}$ we will denote  the density of  the quantity $\mathbf{X}$, 
\be\label{X-bcX}
\mathbf{X}:= \int_{{\cal N}_v} \d{}^{D-2} x \ \bc{X}.
\ee

\subsection{Local First Law at Null Boundary}\label{sec:first-law}
Defining $\bc{P}:={\cal P}/(16\pi G)$ {and $\bc{N}_{AB}:= (16\pi G)^{-1} N_{AB}$}, \eqref{Charges-zero-mode-full} implies,
\be\label{local-first-law}
\boxed{\slashed{\delta} \bc{H}=T_{_{\mathcal{N}}} \delta\bc{S} +{\cal U}^A \delta\bc{J}_A  +\Omega\Theta\delta \bc{P}-{\Omega \Omega^{AB}\delta \bc{N}_{AB}}, \qquad T_{_{\mathcal{N}}}:= -\frac{\Gamma}{4\pi}}
\ee
The above is true at each $v,x^A$ over the null surface and represents the local null boundary  first law. The LHS, unlike the usual first law \eqref{first-law}, is not a complete variation, as the system is describing an open thermodynamic system due to the existence of the expansion and the flux. The above reduces to a usual first law for closed systems when $N_{AB}=0$ or in the non-expanding $\Theta =0$ case.

Note that $\Gamma=-2\kappa+{\cal D}_v\ln (\eta \Omega^{\frac{2}{D-2}})$ where $\kappa$ is the non-affinity parameter (surface gravity) associated the vector field generating the null surface ${\cal N}$ \cite{Adami:2021AAA}. {$-\Gamma/2$ is the local acceleration of an observer for whom $r=0$ is locally the Rindler horizon.} So, $T_{_{\mathcal{N}}}=\frac{\kappa}{2\pi}-\frac{1}{4\pi}{\cal D}_v\ln (\eta \Omega^{\frac{2}{D-2}})$. For non-expanding $\Theta=0$ case where one may put $\eta=1$  or when we have a Killing horizon, $T_{_{\mathcal{N}}}$ equals the usual Unruh/Hawking temperature, \emph{cf.} section \ref{sec:Non-expanding-no-news} for more discussions.

\subsection{Local Extended Gibbs-Duhem  Equation at Null Boundary}\label{sec:Gibbs-Duhem}

Gibbs-Duhem equation \eqref{GD-thermo} is a relation among the thermodynamic charges. Given the expressions for the zero mode charges \eqref{Charges-zero-mode-Integrable} and for the densities in the same notation as in \eqref{X-bcX} we have
\begin{equation}\label{LEGD-eq}
\boxed{\bc{E}=T_{_{\mathcal{N}}} \bc{S} +{\cal U}^A \bc{J}_A }
\end{equation}
The above is an {analogue} of {the}  Gibbs-Duhem equation if $\bc{E}$ is viewed as energy, $\bc{S}$ as entropy and $\bc{J}_A$ as other conserved charges and $\Gamma, {\cal U}^A$ as the respective chemical potentials. This of course manifests the picture we outlined in  section \ref{sec:Thermodynamical phase space}. However, one should note that \eqref{LEGD-eq} is a local equation at the null boundary, unlike its usual thermodynamic counterpart. This equation also holds for non-stationary/non-adiabatic cases when the system is out-of-thermal-equilibrium (OTE). So, we call \eqref{LEGD-eq} `local extended Gibbs-Duhem' (LEGD) equation at the null boundary. 

LEGD equation, like the local first law \eqref{local-first-law}, is a manifestation of  diffeomorphism invariance of the theory. While the explicit expressions for the charges do depend on the theory, we expect \eqref{LEGD-eq} to be universally true for any  diff-invariant  theory of gravity in any dimension. This equation is  on par with the first law of thermodynamics but extends it in two important ways: it is a local equation in $v,x^A$ and holds also for OTE.

Since the integrable parts of the charge are (by definition) independent of the bulk flux $N_{AB}$ and also of ${\cal P}$, the  LEGD equation \eqref{LEGD-eq},  also do not involve $\mathcal{P}$ and  $N_{AB}$. Nonetheless, the chemical potentials in \eqref{LEGD-eq}, $\Gamma$ and ${\cal U}^A$, implicitly depend on $N_{AB}$ and ${\cal P}$ through Raychaudhuri and Damour equations.

\subsection{Local Zeroth Law } \label{sec:Non-expanding-no-news}

Zeroth law in the usual thermodynamics is a statement of thermal equilibrium: as a consequence of the zeroth law, two (sub)systems with the same temperature and chemical potentials are in thermal equilibrium. In the usual thermodynamics flow of charges is proportional to the gradient of associated chemical potentials and hence the absence of such fluxes can be taken as a statement of the zeroth law. In our case, we are dealing with a system  parameterized by chemical potentials 
$\Gamma, {\cal U}^A$ and $\gamma^{AB}$ which are functions of charges $Q_\alpha\in\{\Omega, {\cal P}, {\cal J}_A, N_{AB}\}$. 
This system is not in general in equilibrium but there could be special subsectors which are. The zeroth law is to specify such subsectors. 

Recalling \eqref{GCCE}, flow of charges vanishes on subsystems over which $F_{\partial_v}(\delta_\xi g,g)$ vanishes. On a closely related account, one can show that \cite{progress-ambiguities} this flux has  the same expression as the on-shell variation of the action. Nonetheless, while the charge variation \eqref{surface-charge-01} is invariant under the addition of a total derivative term to the Lagrangian, the Noether charge and hence the flux are not. In particular, upon addition of a boundary Lagrangian $L_{\cal B}=\partial_\mu {\cal B}^\mu$,  the on-shell action variation and hence the flux $F$ are shifted by $\delta {\cal B}^r$. For later convenience, let us call ${\cal B}^r=\bc{G}$. This opens up the possibility of {(partially)} removing the flux by an appropriate boundary term. The question is hence what are the subsectors in the solution phase space for which flux can be removed by an appropriate boundary term.

So, we start with the variation of on-shell action. A direct computation leads to
\begin{equation}\label{Onshell-deltaS-generic}
\begin{split}
    \delta S_{\text{\tiny EH}}|_{\text{\tiny on-shell}}= \frac{1}{16\pi G}\int_{\mathcal{N}} \d v \d{}^{D-2}x \left( 
   \Omega \Theta \delta \mathcal{P}+\Omega \delta \Gamma -{\cal J}_{A}\delta \mathcal{U}^{A} -\Omega N^{AB} \delta\Omega_{AB} 
    \right)=\int \d v  F_{\partial_v} (\delta g;g), 
    \end{split}
\end{equation}
where  $F_{\partial_v}$ may be readily read from \eqref{Non-Integrable-part-Charge-Geometric}. Next, let us add a boundary term to the Lagrangian upon which  $\delta S_{\text{\tiny EH}}|_{\text{\tiny on-shell}}\to \delta S_{\text{\tiny EH}}|_{\text{\tiny on-shell}}+\int_{\cal N} \delta\bc{G}$. As the statement of the  zeroth law we require there exists a $\bc{G}=\bc{G}(\Omega, {\cal P}, {\cal J}_A, N_{AB})$ such that, 
\begin{equation}\label{Zeroth-law---general}
   { \delta \bc{G} = -\bc{S} (\delta T_{_{\mathcal{N}}}-{4G\Theta \delta \bc{P}}) -\bc{J}_{A}\delta \mathcal{U}^{A}  +{\Omega \bc{N}_{AB} \delta\Omega^{AB}}}
\end{equation}
admits non-zero solutions. Integrability condition for the zeroth law \eqref{Zeroth-law---general}  is $\delta (\delta \bc{G})=0$,\footnote{ $\delta(\delta \bc{G})$ is the pre-symplectic two-from \eqref{presymplectic}, therefore \eqref{Zeroth-law---general} implies vanishing symplectic form over the null surface ${\cal N}$.} which yields an equation like $\sum_{\alpha, \beta} C_{\alpha\beta} \delta Q_\alpha \wedge \delta Q_\beta=0$, where $Q_\alpha$ are generic charges {and $C_{\alpha\beta}$ is skew-symmetric}. This equation is satisfied only for $C_{\alpha\beta}=0$. One can immediately see $N_{AB}=0=\delta N_{AB}$ is a necessary (but not sufficient) condition for \eqref{Zeroth-law---general} to have non-trivial solutions.

Before discussing the solutions in more detail, let us note that when \eqref{Zeroth-law---general} is fulfilled the charge $\bc{H}$, which appears in the LHS of the local first law \eqref{local-first-law}, becomes integrable and  we obtain
\begin{equation}
    \boxed{ \bc{H}= \bc{G}+T_{_{\mathcal{N}}} \bc{S} +\mathcal{U}^{A} \bc{J}_{A}}
\end{equation}
Besides $N_{AB}=0$, in terms of $\bc{H}=\bc{H}(\bc{S}, \bc{J}_A,\bc{P})$ local zeroth law implies, 
\begin{equation}\label{zeroth-law-H}
       \boxed{T_{_{\mathcal{N}}}= \frac{\delta \bc{H}}{\delta \bc{S}} \,  , \qquad \mathcal{U}^A =  \frac{\delta \bc{H}}{\delta \bc{J}_{A}} \, , \qquad {\cal D}_v\bc{S} =\bc{S} \Theta= \frac{1}{4G}\ \frac{\delta \bc{H}}{\delta \bc{P}}}
\end{equation}
where the last equation may be seen as the \emph{equation of state}. For the special case of $\Theta =0$, one simply deduces that $\bc{H}$ does not depend on $\bc{P}$.  Eq.\eqref{Zeroth-law---general} ensures that total energy and  angular momentum are conserved {on-shell},  $\frac{\d {} }{\d v} \mathbf{H}= \frac{\d {} }{\d v} \mathbf{J}_A=0$, where $\frac{\d {} }{\d v}\mathbf{X}:= \int_{{\cal N}_v} \d{}^{D-2} x \ \partial_v\bc{X}$.  Total entropy, on the other hand, is not conserved as $\frac{\d {} }{\d v} \mathbf{S}= \int_{\mathcal{N}_v} \d{}^{D-2}x\ \Theta \, \bc{S}$; $\frac{\d {} }{\d v}\mathbf{S}$ is zero only when expansion vanishes, $\Theta=0$.

\paragraph{Generic $\Theta\neq 0$ case.} The zeroth law requires $N_{AB}=0$ for which 
\eqref{RD-EoM-2} reduce to
\be\label{RD-EoM--N=0}
T_{_{\mathcal{N}}}=- 4G {\cal D}_v \bc{P},\qquad {\cal D}_v \left[\bc{J}_A+4G \bar\nabla_A(\bc{S}\bc{P})\right]=0.
\ee
The above imply that zeroth law  \eqref{zeroth-law-H} is satisfied for any $\bc{H}=\bc{H}(\bc{S},\bc{P}, \bc{J}_A)$, when   $\bc{S}, \bc{P}$ and $\bc{J}_A$ have the following basic Poisson brackets \cite{Adami:2021AAA}
\be\label{Charge-brackets-Theta-neq0}
\begin{split}
  &\{\bc{S}(x,v), \bc{P}(y,v)\}=\frac{1}{4G}\delta^{D-2}(x-y),\quad  \{\bc{S}(x,v), \bc{S}(y,v)\}= \{\bc{P}(x,v), \bc{P}(y,v)\}=0,\\
&\{\bc{S}(x,v), \bc{J}_A(y,v)\}= \bc{S}(y,v)\frac{\partial}{\partial x^{A}}\delta^{D-2}(x-y), \\
&\{\bc{P}(x,v), \bc{J}_A(y,v)\}= \left(\bc{P}(y,v)\frac{\partial}{\partial x^{A}}+ \bc{P}(x,v)\frac{\partial}{\partial y^{A}}\right)\delta^{D-2}(x-y), \\
 &\{\bc{J}_A(x,v), \bc{J}_B(y,v)\}=\frac{1}{16\pi G}\left(\bc{J}_A(y,v)\frac{\partial}{\partial x^{B}}-\bc{J}_B(x,v)\frac{\partial}{\partial y^{A}}\right)\delta^{D-2}(x-y)
\end{split}
\ee
and ${\partial_v} \bc{X}=\{\bc{H}, \bc{X}\}.$  That is,  $\bc{H}$ is the Hamiltonian over this phase space and \eqref{zeroth-law-H} do not impose any restrictions on $\bc{H}$ which is a scalar over ${\cal N}$.

\paragraph{$\Theta=0$ case.} In this case trace of {the} extrinsic curvature of the null surface ${\cal N}$ vanishes,   hence {it is} an extremal null surface.  Vanishing of {the} expansion $\Theta $ has some important consequences. (1) Raychaudhuri equation implies $N_{AB}=0$. So, again we arrive at the vanishing flux; (2) $\eta$ drops out from the charge variation \eqref{surface-charge-01}.  (3) We lose one tower of the charge ${\cal P}$, and the associated symmetry generator becomes a pure gauge. (4) We may fix the ${\eta }=1$ gauge which yields $W=2{\cal D}_v T$. We hence remain with $T, Y^A$ generators which form Diff(${\cal N})$ symmetry algebra. (5) EoM \eqref{RD-EoM} reduces to ${\cal D}_v \bc{J}_A=\bc{S}\partial_A T_{_{\mathcal{N}}}$ and  ${\cal D}_v\bc{S}=0$, {which may be viewed as equations for spatial derivatives of the chemical potentials.}

Local zeroth law  \eqref{zeroth-law-H} is satisfied by any scalar Hamiltonian $\bc{H}=\bc{H}(\bc{S}, \bc{J}_A)$, together with basic Poisson brackets  \eqref{Charge-brackets-Theta-neq0} but with $\bc{P}$ dropped \cite{Adami:2021AAA} and again with $\partial_v \bc{X}=\{\bc{H}, \bc{X}\}$.

\paragraph{Closing remarks:}  (1) \tcb{Local} zeroth law \eqref{zeroth-law-H} is just defining the Poisson bracket structure over our charge space and existence of  Hamiltonian dynamics, {but} does not specify a Hamiltonian. (2) Choice of Hamiltonian fixes a boundary Lagrangian and the boundary dynamical equations which in turn specifies local dynamics of charges on the null boundary ${\cal N}$. 
(3) In analogy with isolated horizon \cite{Ashtekar:2000sz} of black holes, if the zeroth law holds the null surface may be called an `isolated null surface'.\footnote{One may 
show by direct computation that, upon zeroth law in the $\Theta=0$ sector, ${\cal D}_v \bc{J}_A=\bc{S}\partial_A T_{_{\mathcal{N}}}$ and ${\cal D}_v\bc{S}=0$ and also $\mathcal{D}_v \bc{E} =- \partial_v \bc{G}, \mathcal{D}_v \bc{H}= - \partial_A (\mathcal{U}^A \bc{G})$.} (4) Our zeroth law is a weaker condition than stationarity as $\partial_v$ of the chemical potentials need not vanish. (5) The usual zeroth law of black hole mechanics (for Killing horizons) that ${\cal U}^A$ and $\Gamma$ are constants over the horizon (our null boundary ${\cal N}$) is a very special case which obeys our local zeroth law. For {the} stationary asymptotic flat black hole solutions to {the} vacuum Einstein gravity, {i.e.} the Myers-Perry solutions, we {get} $\bc{E}= \left(\frac{D-3}{D-2}\right) \bc{H}$, and we have the usual Smarr formula.

\section{Outlook}\label{sec:discussion}

Building upon the analyses of \cite{Adami:2020amw, Grumiller:2020vvv, Adami:2020ugu, Adami:2021sko} and in particular \cite{Adami:2021AAA}, we established that the solution phase space around an arbitrary null surface in pure $D$ dimensional Einstein gravity naturally admits a thermodynamical description and the charges and  corresponding chemical potentials form a `thermodynamical phase space'. The laws of thermodynamics are all local equations over the $D-1$ dimensional null surface ${\cal N}$ and our analysis does not fix the boundary dynamics, boundary Hamiltonian, which may still be chosen.

As discussed, the zeroth law necessitates vanishing of the flux of bulk gravitons through ${\cal N}$ and establishes basic Poisson brackets on the thermodynamic phase space. The same condition, absence of $N_{AB}$, has been discussed  as the condition for {the} existence of  a  slicing in the solution phase space in which the charges become integrable \cite{Grumiller:2020vvv, Adami:2021sko, Adami:2021AAA}. The physics of change of phase space slicing and null surface thermodynamics developed here is an interesting direction to for further investigations.

{In this work we focused on the zeroth and first laws and the Gibbs-Duhem equation. However, the second law  of thermodynamics is an important part of any thermodynamic description. In the black hole thermodynamics, there is the ``generalized second law'' stating that the the sum of entropies of black hole and the outside do not decrease, see \cite{Wall:2018ydq} and references therein. A simplified version of the second law is Hawking's area theorem that in black hole merger processes the area of horizon does not decrease  which uses Rachyaudhuri equation (focusing theorem) and null energy condition for the matter fields, see e.g. \cite{Chrusciel:2000cu}. In our setting, one may look for a local version of the second law, recalling that flux of gravitons can only move through the null boundary to the ``inside region'' and nothing comes out. See in particular the analysis in section 8 of \cite{Adami:2021AAA}.  As the first relevant step towards a local second law we have worked through ``null surface focusing theorem'' in the appendix. Further analysis and discussion on this very important issue is left for future work. }

Our analysis is based on covariant phase space formulation and hence readily generalizes to any diffeomorphism invariant theory in any dimension. Given all the previous literature, especially \cite{Iyer:1994ys}, it is {reasonably} certain that the same `local' thermodynamical description with  {exactly} the same equations should also hold for this generic case. In other words, diffeomorphism invariance yields the local thermodynamical picture. One may do the reverse and show that the thermodynamical description results in diffeomorphism invariance. 
We should emphasize that this is already weaker than Einstein's equivalence principle and non-minimal coupling and generic modified gravity theories follow the same analysis. The connection between thermodynamics and gravity is nothing new, e.g. see \cite{Jacobson:1995ab}. {Our analysis here, derives the local thermodynamics relation which is assumed in \cite{Jacobson:1995ab} from first principles and do not require the null boundary to have any extra properties.} {Our approach here, in contrast to the usual viewpoint  puts the emphasis on the boundary phase space, rather than the bulk graviton modes. In our local thermodynamics description, the latter appear as the flux (news) through the boundary. We hope our new ``boundary based'' viewpoint and framework adds new shed on thermodynamics/gravity relations and can be pushed to the quantum level.} 

The local thermodynamical description in its basics and general ideas reminds one of the membrane paradigm \cite{Price:1986yy, Thorne:1986iy, Parikh:1998mg}. It is interesting to relate the two  more systematically. The first steps in this direction {were} outlined in \cite{Grumiller:2018scv}. The interesting question is whether the membrane picture restricts the boundary Hamiltonian.

Among other things, our analysis here very clearly shows how the boundary and bulk degrees of freedom interact and that the boundary phase space admits the local thermodynamical description. This is expected to be very relevant for the black hole microstate problem in that the boundary d.o.f and quantization thereof can account for the microstates, whereas the interactions with bulk modes would be relevant for the information loss problem. Our analysis permits a semi-classical setting in which boundary phase space is quantized while bulk modes are treated (semi)classically and hence potentially gives a better handle on both microstate and information loss problems.

\paragraph{Acknowledgement.}
We would like to especially thank Daniel Grumiller and Celine Zwikel for long term collaborations and  many fruitful discussions which were crucial in developing the ideas and analysis discussed here. We thank Glenn Barnich for correspondence and Mohammad Hassan Vahidinia for the useful discussions. MMShJ acknowledges SarAmadan grant No. ISEF/M/400122. The work of VT is partially supported by IPM funds. {The work of HA was supported by the NSFC Grant No. 12150410311.}

\appendix
\section{Null Boundary  Focusing Theorem} \label{sec:Focusing}

From Raychaudhuri equation \eqref{EoM-Raychaudhuri} one learns that,
\be\label{Raychaudhri-1}
{\cal D}_v \Theta -\kappa \Theta +\frac{1}{D-2}\Theta ^2 \leq 0, \qquad \kappa:=-\frac{\Gamma}{2}+\frac12{\cal D}_v\ln (\eta \Omega^{\frac{2}{D-2}}), 
\ee
In terms of variable 
\be\label{XX}
X(v):=\exp{\left(\int_{v_0} \kappa\right)}, \qquad {\cal D}_v X-\kappa X=0,\quad X(v_0)=1,
\ee
 \textit{assuming $\Theta (v)\neq 0$}, \eqref{Raychaudhri-1} implies
\be\label{Focusing-1}
\Theta \leq \frac{\Theta ^0 X(v)}{1+\frac{\Theta ^0}{D-2}\int_{v_0}  X(v)},
\ee
where $\Theta ^0=\Theta (v=v_0)$ and without loss of generality we have taken $v_0$ such that the equality is saturated. Since $X(v)\geq 0$, then $\int_{v_0}  X(v)$ is a growing function of $v$. If $\Theta ^0<0$ (that is if we start in a contracting phase) then there will always be a ``trapping time'' $v_1>v_0$ where $\int_{v_0}^{v_1}  X(v)=-\frac{1}{\Theta ^0 (D-2)}$, and $\Theta (v_1)\to -\infty$.

If at some $v$, $\hat{v}_0$, $\Theta =0$ then its derivative should be non-positive at that point ${\cal D}_v\Theta \leq 0$. For the case of equality we have a non-expanding case and the $N_{AB}$ should also vanish and if ${\cal D}_v\Theta <0$ then at $\hat{v}_0+\delta v$, $\Theta <0$ and again \eqref{Focusing-1} implies existence of a trapping time. We should stress that all the above analysis is \textit{local} on the codimension 2, constant $v,r$ surfaces and all quantities are functions of $x^A$.

In the absence of bulk modes, $N_{AB}=0$, the above inequality is replaced with equality. In this case \eqref{Focusing-1} shows internal null boundary dynamics which is of course due to gravity effects. In other words, gravity dynamics is relating thermodynamical sector of the solution phase space to the other two parts, the $\eta$ part and the bulk modes and  this dynamics is featured in the focusing equation.

\bibliographystyle{fullsort.bst}

\providecommand{\href}[2]{#2}\begingroup\raggedright\endgroup

\end{document}